\documentclass[10pt,aps,twocolumn,prb,superscriptaddress,bibnotes,amsmath,amsfonts,floatfix,longbibliography,a4paper]{revtex4-2}
\pdfoutput=1
\usepackage{graphicx}
\usepackage{multirow}

\usepackage{lmodern}
\usepackage{textcomp}

\IfFileExists{microtype.sty}{\usepackage{microtype}}{\relax}
\usepackage[pdfusetitle]{hyperref}

\def\rmd{{\rm d}}
\def\rmi{{\rm i}}

\def\im{\mathop{\rm Im}\nolimits}
\def\tr{\mathop{\rm Tr}\nolimits}

\def\muB{\textmu\textsubscript{B}}

\let\vec=\mathbf

\def\up{\uparrow}
\def\dn{\downarrow}

\def\m{\phantom{-}}
\def\0{\phantom{0}}

\def\figfrac{0.95}

\usepackage[dvipsnames]{xcolor}

\let\change=\relax

\begin{document} 

\title{Electronic structure and magnetism in UGa\textsubscript{2}:
  DFT+DMFT approach}

\author{Banhi Chatterjee}
\affiliation{Institute of Physics (FZU), Czech Academy of Sciences, Na
  Slovance 2, 182 21 Prague, Czech Republic}
\affiliation{Jo\v{z}ef Stefan Institute, Jamova 39, SI-1000 Ljubljana,
  Slovenia}
\author{Jind\v{r}ich  Koloren{\v{c}}}
\email{kolorenc@fzu.cz}
\affiliation{Institute of Physics (FZU), Czech Academy of Sciences, Na
  Slovance 2, 182 21 Prague, Czech Republic}

\hypersetup{
pdfauthor = {Banhi Chatterjee (https://orcid.org/0000-0003-3032-6717),
  Jindrich Kolorenc (https://orcid.org/0000-0003-2627-8302)}
}

\date{\today}

\begin{abstract}
The debate whether uranium 5f electrons are closer to being localized
or itinerant in the ferromagnetic compound UGa\textsubscript{2} is not
yet fully settled. The experimentally determined magnetic moments are
large, approximately 3~\muB, suggesting the localized character of the
5f~electrons. In the same time, one can identify signs of itinerant as
well as localized behavior in various spectroscopic observations. The
band theory, employing local exchange-correlation functionals, is
biased toward itinerant 5f~states and severely underestimates the
moments. Using material-specific dynamical mean-field theory (DMFT),
we probe how a less approximate description of electron-electron
correlations improves the picture. We present two variants of the
theory: starting either from spin-restricted (LDA) or spin-polarized
(LSDA) band structure. We show that the L(S)DA+DMFT method can
accurately describe the magnetic moments in UGa\textsubscript{2} as
long as the exchange interaction between the uranium 6d and 5f
electrons is preserved by a judicious choice of the spin-polarized
double-counting correction. We discuss the computed electronic
structure in relation to photoemission experiments and show how
the correlations reduce the Sommerfeld coefficient of the electronic
specific heat by shifting the 5f states slightly away from the Fermi
level.
\end{abstract}

\maketitle


\section{Introduction}
\label{sec:intro}

The 5f electrons in actinides and their compounds can be either
itinerant and participating in chemical bonds, or localized and not
contributing to cohesion. A transition akin to Mott metal--insulator
transition occurs in elemental actinide metals between Pu and Am
\cite{moore2009}.  Although elemental uranium has itinerant 5f electrons,
its compounds display both types of 5f states. A traditional way of
classifying uranium compounds is by placing them in the Hill plot that
relates the critical temperature (magnetic or superconducting) to
the nearest neighbor U--U spacing \cite{hill1970}. Small U--U
distances favor superconducting behavior at low temperatures,
whereas long-range magnetic order takes place at spacings greater
than the so-called Hill limit ($3.5$~\AA).

\begin{figure}
\hbox to\linewidth{\hfill%
\includegraphics[width=0.86\linewidth]{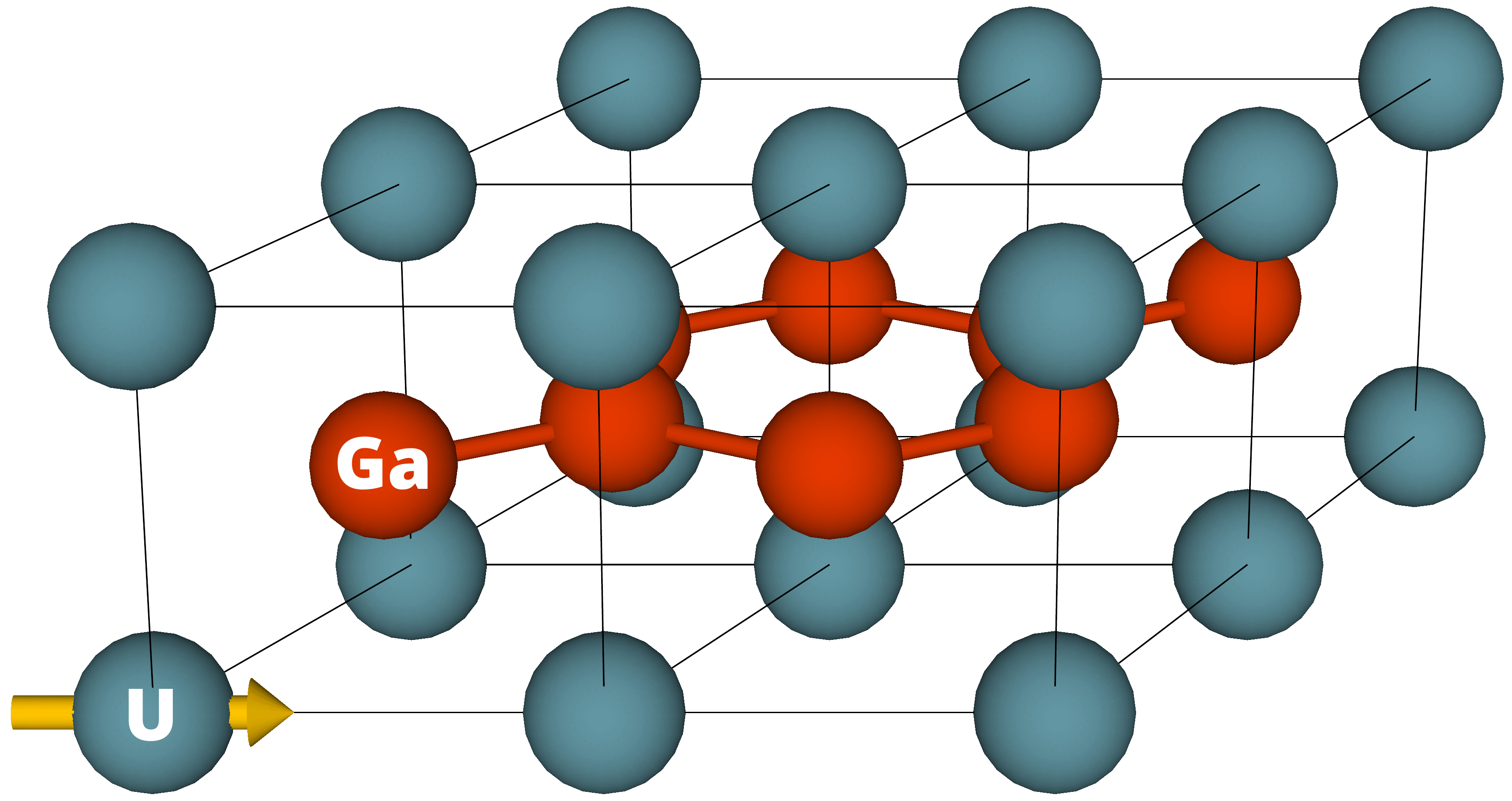}%
\hfill}
\caption{\label{fig:latticestruct} The hexagonal lattice of
  UGa${}_2$ with uranium atoms shown in blue and gallium atoms in
  red. The uranium magnetic moments (arrow) are aligned along the
  [100] direction.}
\end{figure}

In UGa\textsubscript{2}, an intermetallic binary compound with a
hexagonal AlB\textsubscript{2} structure (space group P6/mmm,
Fig.~\ref{fig:latticestruct}), the Ga atoms effectively separate the
uranium atoms, increasing the U--U distance to $4.0$~\AA, that is,
above the Hill limit. Accordingly, the compound exhibits ferromagnetic
order below $T_\text{C}=125$~K with the easy magnetization axis along the [100]
direction. Experimental observations establish magnetic moments of
approximately 3~\muB\ per U atom in the ferromagnetic phase, using
magnetization measurements \cite{andreev1978,kolomiets2015} as well as
neutron diffraction \cite{lawson1985,ballou1982u4+}. UGa${}_2$ thus
exhibits moments and ordering temperature that are larger than typical
for ferromagnetic uranium intermetallics \cite{handbook}, which
indicates localized 5f electrons. The magnetic behavior can indeed be
accurately reproduced by a fully local crystal-field model
corresponding to the 5f\textsuperscript{3} configuration of the U ion
\cite{radwanski1995}. In addition, the observed Sommerfeld coefficient
$\gamma=11$~mJ/mol$\cdot$K\textsuperscript{2} \cite{honma2000} is not
much enhanced compared to the analogous compound without 5f electrons
-- LaGa\textsubscript{2}, displaying
$\gamma=4$~mJ/mol$\cdot$K\textsuperscript{2} \cite{fujimaki1992}, which
testifies against a high density of electronic states at the Fermi
level in UGa\textsubscript{2}, again favoring the localized picture of
the 5f electrons. The spectroscopic evidence, on the other hand, is
not conclusive about the nature of the 5f states since one can
identify spectral features characteristic to localized electrons as
well as features typical to itinerant electrons
\cite{gouder2001b,fujimori2019,kolomiets2021}. Similarly, the Fermi
surface probed by the de Haas--van Alphen effect is not compatible with
full 5f localization \cite{honma2000}.

The large spin-orbit coupling (SOC), the crystal-field splitting, and
the Coulomb interaction between the 5f electrons influence the
magnetic moments in a non-trivial manner. This complexity contributes
to the fact that the electronic structure of UGa\textsubscript{2} is
not yet satisfactorily understood. The first-principles band theory
based on semi-local approximations to the density-functional
theory (DFT) severely underestimates the moments, yielding about
0.6~\muB\ per uranium atom \cite{divis1996,chatterjee2020}.
The correlated band theory incorporating an on-site Hubbard
interaction term, DFT+$U$, can successfully model the magnetically
ordered states, particularly in insulating compounds with localized 5f
electrons \cite{anisimov1997b,suzuki2013,qiu2020effective}.  In
UGa\textsubscript{2}, it enhances the magnetic moments up to 2.8~\muB\
but the spectroscopic results are not reproduced very well
\cite{chatterjee2020,antonov2003electronic}.

{\change The DFT+$U$ method is a static mean-field approximation and
  as such it} cannot
account for the multi-reference character of the 5f shell nor
for dynamical many-body effects. These limitations are lifted when DFT
is combined with the dynamical mean-field theory (DMFT)
\cite{lichtenstein1998,kotliar2006}, which accurately models both
itinerant and localized electrons. In this paper, we investigate how
the theoretical description of the magnetism and of the electronic
structure of UGa\textsubscript{2} improves when the DFT+DMFT is
applied. We estimate and discuss the effects of the 6d--5f exchange
interactions on the 5f magnetic moments, and compare the computed
spectral properties with the experimental valence-band photoemission
spectra (PES). We also discuss technical matters pertaining to
spin-polarized DFT+DMFT solutions.


\section{Method}
\label{sec:method}

{\change The DFT+DMFT method improves upon DFT+$U$ by replacing the
static mean-field potential, approximating the Coulomb interaction
among the uranium 5f~electrons, with} an energy-dependent (dynamical)
potential (self-energy) \cite{georges1996,lichtenstein1998}. This self-energy
is computed by solving an auxiliary impurity model {\change -- a many-body
problem}, for which we employ 
the exact diagonalization. We present two variants of the
theory differentiated by the self-energy being inserted (a) into the
spin-restricted LDA band structure (we call this method LDA+DMFT), and
(b) into the ferromagnetic LSDA band structure (we refer to this
variant as to LSDA+DMFT). A similar comparison of spin-restricted and
spin-polarized parent band structures was performed for ferromagnetic
nickel in \cite{katsnelson2002a}.


\subsection{General formalism}
\label{sec:methodGeneral}

We start with determination of the first-principles band structure by
means of the WIEN2k code \cite{wien2k} using parameters listed in
Appendix~\ref{app:DFT_params}. Scalar relativistic effects as well as
the spin-orbit coupling are included in these WIEN2k
calculations {\change \cite{koelling1977}}. Afterwards, the relevant
valence bands are represented
by a tight-binding Hamiltonian in the basis of the maximally-localized
Wannier functions \cite{kunes2010,mostofi2008}. This Hamiltonian is
then used as the parent band structure for the DMFT calculations.

In each iteration of the DMFT self-consistency cycle, the local
electronic structure around one shell of the uranium 5f Wannier
functions is mapped onto a non-interacting impurity model
(Appendix~\ref{app:bath_discretization}),
\begin{multline}
\label{nonintsiam}
\hat H_{\rm imp}=
\sum_{mm'\sigma\sigma'}
\bigl[\mathbb{H}_\text{loc}\bigr]_{mm'}^{\sigma\sigma'}
 \hat f_{m\sigma}^{\dag}\hat f_{m'\sigma'}
+\sum_{J}\epsilon_J\hat b_J^{\dag}\hat b_J\\
+\sum_{m\sigma J}\Bigl(V_{J m\sigma}\hat f_{m\sigma}^{\dag}\hat b_J
  +V^*_{J m\sigma}\hat b_J^{\dag}\hat f_{m\sigma}\Bigr)\,,
\end{multline}
where $\hat f_{m\sigma}^\dag$ creates an electron in the 5f shell with
magnetic quantum number $m$ and spin projection $\sigma\in\{-1/2,1/2\}$ 
(eigenvalues of $\hat s_z$). The first term in Eq.~\eqref{nonintsiam}
corresponds to the local Hamiltonian, which describes the
5f~shell. It can be decomposed as
\begin{multline}
\mathbb{H}_\text{loc}= \epsilon_{f} \hat{I}+\zeta \hat{\mathbf {l}} \cdot \hat{\mathbf {s}}
-\hat{\mathbf{s}}\cdot\boldsymbol{\Delta}_X\\
+B_{20}\hat O_{20} + B_{40}\hat O_{40} + B_{60}\hat O_{60}
+B_{66}\hat O_{66}\,,
\label{eq:HlocDecomp}
\end{multline}
where $\epsilon_{f}$ is the energy of the 5f level, $\zeta$ is the
strength of the SOC, $\boldsymbol{\Delta}_X$ gives the exchange
splitting, and $\hat O_{kq}$ and $B_{kq}$ are Stevens operators and
the corresponding parameters that characterize the D\textsubscript{6h}
crystal-field potential at the uranium site in
UGa\textsubscript{2}. In general, the parameters $B_{kq}$ can be spin
dependent, which we briefly discuss at the end of
Appendix~\ref{app:Hloc}. Note that the decomposition introduced in
Eq.~\eqref{eq:HlocDecomp} is only used for the analysis of
$\mathbb{H}_\text{loc}$ and has no influence on the DMFT calculations
and results.

The second term in Eq.~\eqref{nonintsiam} corresponds to an effective 
medium usually referred to as the bath, with which the 5f~shell interacts.
The operator $\hat b_{J}^\dag$ creates an electron in this effective medium.
The last term in Eq.~\eqref{nonintsiam} accounts for the hybridization
of the 5f shell with the 
bath. In our calculations, the off-diagonal hybridization induced by
the non-commutativity of the hexagonal symmetry with the SOC is
fully taken into account. The crystal-field splitting of the 5f states
is partly due to the crystal-field potential contained in
$\mathbb{H}_\text{loc}$ and partly due to the hybridization.

The full interacting impurity model, in which the self-energy is
computed, is given by
\begin{equation}
 \hat{H}_\text{imp}^\text{DMFT} = \hat{H}_\text{imp} +\hat{U},
  \label{intsiam}
\end{equation}
where $\hat{H}_\text{imp}$ is the non-interacting one-electron part
shown in Eq.~\eqref{nonintsiam} and $\hat{U}$ is the Coulomb repulsion
among the 5f electrons,
\begin{multline}
\hat{U} = \frac{1}{2}\sum_{\substack {m m'm''\\m'''\sigma\sigma'}}
 U_{m m'm''m'''} \hat{f}_{m\sigma}^\dag \hat{f}_{m' \sigma '}^\dag
 \hat{f}_{m '''\sigma '} \hat{f}_{m'' \sigma} \\[-1em]
-\sum _{m\sigma}( U_{H} - \sigma U_{X})
 \hat{f}_{m\sigma}^\dag \hat{f}_{m \sigma}\,,
\label{coulumbU}
\end{multline}
where $U_{mm'm''m'''}$ is considered in its full spherically symmetric
form parametrized by four Slater integrals $F_{0}=2.0$~eV,
$F_{2}=7.09$~eV, $F_{4}=4.60$~eV, and $F_{6}=3.36$~eV, which
correspond to Coulomb $U=2.0$~eV and Hund $J=0.59$~eV. The first
integral, $F_0$, is at the upper limit, beyond which the 5f peak in
the occupied LSDA+$U$ density of states moves too far from the
Fermi level to be compatible with the valence-band photoemission
spectra \cite{gouder2001b,chatterjee2020}. The other three
parameters ($F_2$, $F_4$, $F_6$) correspond to the atomic
Hartree--Fock values calculated for the U$^{3+}$ ion (5f$^{3}$
configuration) and then reduced to 80\% to mimic screening
\cite{cowanCode,ogasawara1991}. Note that the unscreened ionic $F_k$
values yield Hund $J=0.79$~eV, which can be considered as
the maximal value for the uranium 5f$^{3}$ systems.

The second term in Eq.~\eqref{coulumbU} is the double-counting
correction introduced to remove the static mean-field approximation of
the 5f--5f Coulomb interaction that is incorporated in the DFT band
structure. We assume the double-counting correction to be spherically
symmetric (neither $U_H$ nor $U_X$ depends on the magnetic quantum
number $m$), with $U_X=0$ for the LDA band structure and $U_X\neq0$
for the LSDA band structure. The numerical values of $U_H$
and $U_X$ are discussed later {\change in Secs.~\ref{sec:methodLDADMFT}
and~\ref{sec:methodLSDADMFT}.}

The impurity model, Eq.~\eqref{intsiam}, is solved using the exact
diagonalization (Lanczos) method \cite{meyer1989,liebsch2012} as
implemented in our in-house code \cite{kolorenc2015b}. The size of the
models, which can be solved by this method, is limited due to
unfavorable scaling of the computational demands. The impurity
models employed in this paper consist of 14 spinorbitals corresponding
to the 5f shell and another 42 spinorbitals representing the bath. Of
the bath states, $N_b^<=14$ orbitals have $\epsilon_J$ below the Fermi
level (they are nominally occupied) and $N_b^>=28$ orbitals have
$\epsilon_J$ above the Fermi level (they are nominally empty). Even
these small models are too demanding unless we turn to
a reduction of the many-body basis inspired by the work of Gunnarsson
and Sch\"onhammer \cite{gunnarsson1983,kolorenc2015b}. A cutoff $M$ is
introduced for each $N$-electron Hilbert space~$\mathcal{H}_N$, and
the diagonalization is performed only in a subspace
\begin{equation}
\label{eq:EDtruncHilbert}
\mathcal{H}_N^{(M)}=\bigl\{
|f^{N-N_b^<-n+m}\,b^n\,\text{\underline{$b$}}^m\rangle,\,
0\leq m+n\leq M\bigr\}\,.
\end{equation}
In this notation, $f^{N-N_b^<-n+m}$ indicates $N-N_b^<-n+m$ electrons
in the uranium 5f shell, $b^n$ indicates $n$ electrons in the bath orbitals
above the Fermi level, and $\text{\underline{$b$}}^m$ means $m$
holes in the bath orbitals below the Fermi level. We use $M=2$ for the
cutoff. {\change The convergence with respect to $M$ is discussed in
Appendix~\ref{app:Mconv}.}

The impurity solver yields a self-energy
$[\hat\Sigma(z)]_{mm'}^{\sigma\sigma'}$ acting in the subspace of 5f 
spinorbitals, which enters the Dyson equation for the local Green's
function $\hat G(z)$, 
\begin{equation}
\label{eq:localGFcrystal}
\hat G(z)=\frac1{\mathcal N}\sum_{\mathbf k}\bigl[ z\hat I - \hat H_\mathbf{k}
 -\hat\Sigma(z)\bigr]^{-1}\,,
\end{equation}
where $\mathcal N$ is the number of k points in the Brillouin zone (4096 in our
calculations) and $\hat H_\mathbf{k}$ is the tight-binding
Hamiltonian. The local Green's function determines an updated impurity
model (Appendix~\ref{app:bath_discretization}), concluding one
iteration of the DMFT cycle.

After the DMFT self-consistency is reached, the occupation matrix of
the 5f states is evaluated from the 5f block of the local Green's
function,
\begin{equation}
\hat n_f= \int_{-\infty}^{E_\text{F}}
\hat A_f(\epsilon)\,\rmd\epsilon\,,\
\hat A_f(\epsilon)
  =-\frac{1}{\pi} \im \hat G_f(\epsilon+\rmi0)\,,
\end{equation}
where the integral runs over all occupied states up to the Fermi energy
$E_\text{F}$. Knowing the occupation matrix, we can calculate the 5f electron
occupation as well as spin and orbital moments as averages of the
corresponding operators,
\begin{equation}
 n_{f} = \tr (\hat{n}_f)\quad\text{and}\quad
 \langle O \rangle =  \tr (\hat{O}\, \hat{n}_f)\,.
 \label{eq:operator_averages}
\end{equation}
Finally, the Sommerfeld coefficient of the electronic specific heat
$\gamma$ is evaluated using the Fermi-liquid formula,
\begin{equation}
\gamma=\frac{\pi k_\text{B}^2}{3}\biggl[
  \frac{g_f(E_\text{F})}{Z_f}
 +g_{spd}(E_\text{F})\biggr]\,,
 \label{sommerfeld}
\end{equation}
where $g_f(E_\text{F})=\tr\bigl[\hat
A_f(E_\text{F})\bigr]$ is the density of 5f states at the 
Fermi energy, $g_{spd}(E_\text{F})$ is the
density of all other states at the Fermi energy, and $Z_f<1$ is
the average quasiparticle weight in the 5f bands that is estimated
from the DMFT self-energy as suggested in \cite{pourovskii2007},
\begin{equation}
\frac1{Z_f}=
\tr\biggl[ \frac{\hat A_f(E_\text{F})}{g_f(E_\text{F})}
\biggl(\hat I-\frac{\rmd\hat\Sigma(\epsilon+\rmi0)}{\rmd\epsilon}\biggr)
\Bigr|_{\epsilon=E_\text{F}}
\biggr]\,.
\end{equation}

All DMFT calculations presented in this paper are performed at
temperature $T=0$~K in order to obtain the ferromagnetic state with
saturated magnetic moments.


\subsection{Choice of the tight-binding model}

\begin{table*}
\caption{\label{loc_ham}Characteristics of several tight-binding models
  derived from the DFT band structure. All models contain gallium 4s
  and 4p orbitals, the included uranium orbitals are listed in the
  first column. The quantities $\zeta$, $\epsilon_f$ and
  $\Delta_{X}$ are shown in eV, the crystal-field parameters $B_{kq}$ in
  meV.}
\begin{ruledtabular}
\begin{tabular}{l c c c c c c c c c c c c c c c c }
\multirow{2}{*}{model} & \multicolumn{9}{c}{orbital occupations} &
\multicolumn{7}{c}{local Hamiltonian $\mathbb{H}_\text{loc}$} \\
\cline{2-10}\cline{11-17}
& U 5f  & U 5f$\up$ & U 5f$\dn$ & U 6d$\up$ & U 6d$\dn$ & U 7s& U 7p& Ga 4s& Ga 4p
&$\zeta$ & $\epsilon_f$ & $\Delta_{X}$
& $B_{20}$ &$B_{40}$ &$B_{60}$ &$B_{66}$ \\
\hline
\multicolumn{8}{l}{nonmagnetic solution}\\
d,f    &2.79&    &    &0.94&0.94& -- & -- &1.51&2.18&0.248&0.634&     0 & $-0.72$&$-0.14$&$\m0.00$&$-0.19$\\
s,d,f  &2.77&    &    &0.95&0.95&0.35& -- &1.50&2.03&0.248&0.639&     0 & $-0.69$&$-0.12$&$ -0.01$&$-0.16$\\
s,p,d,f&2.72&    &    &1.03&1.03&0.76&0.74&1.39&1.50&0.251&0.679&     0 & $-2.83$&$-0.01$&$\m0.00$&$-0.06$\\[1ex]
\multicolumn{8}{l}{ferromagnetic solution [001]}\\
s,d,f  &2.77&2.41&0.37&1.00&0.87&0.35& -- &1.49&2.02&0.246&0.926&0.972&$\m5.98$&$ -0.11$&$ -0.01$&$-0.16$\\
s,p,d,f&2.72&2.37&0.36&1.08&0.96&0.76&0.74&1.39&1.49&0.249&0.968&0.980&$\m3.71$&$\m0.01$&$\m0.00$&$-0.05$\\[1ex]
\multicolumn{8}{l}{ferromagnetic solution [210]}\\
s,p,d,f&2.72&2.34&0.38&1.08&0.95&0.76&0.74&1.34&1.54&0.248&0.956&0.956&$\m3.92$&$\m0.03$&$\m0.00$&$-0.04$\\
\end{tabular}
\end{ruledtabular}
\end{table*}

We investigated several tight-binding models $\hat H_\mathbf{k}$ of
increasing size. As the minimal model, we considered one that contains
gallium 4s and 4p, and uranium 5f and 6d states. Then we included
uranium 7s and finally also 7p states. Various characteristics of
these models are listed in Table~\ref{loc_ham}. Although the uranium
7p states are relatively high above the Fermi level, their inclusion
makes a sizable difference, in particular to the crystal-field
parameters in $\mathbb{H}_\text{loc}$ and to the filling of the
gallium states.

On the top of that, we found that the LDA+DMFT calculations without the U
7p states converge to the out-of-plane [001] ferromagnetic state, whereas
the calculations with the U 7p states predict an in-plane
ferromagnetic state. Since the experiments determine
UGa\textsubscript{2} to be an in-plane ferromagnet
\cite{andreev1978,kolomiets2015}, all results presented in the
following sections were obtained in the tight-binding models that
include uranium 7s and 7p states.


\subsection{LDA+DMFT}
\label{sec:methodLDADMFT}

When the parent band structure is spin-restricted (LDA), we induce the
ferromagnetic solution by introducing a small symmetry-breaking
magnetic field into the impurity model, Eq.~\eqref{nonintsiam}, in the
first few iterations of the DMFT self-consistency cycle. Afterwards,
this field is removed again. Since we do not implement any charge
self-consistency, the tight-binding Hamiltonian $\hat H_{\vec k}$
remains unchanged during the whole LDA+DMFT cycle and the spin (and
orbital) polarization is introduced only by means of the polarized
self-energy applied to the 5f states. This method very likely results in an
underestimated spin polarization of the 6d bands. Moreover, the local
Hamiltonian $\mathbb H_\text{loc}$ stays non-polarized as demonstrated
in Appendix~\ref{app:Hloc}, that is, no exchange
field~$\boldsymbol{\Delta}_X$ is induced in $\mathbb H_\text{loc}$ by
the polarized self-energy. Nevertheless, there should be some exchange field
present in $\mathbb H_\text{loc}$ due to the partially filled and
partially polarized 6d bands, and 
neglecting this exchange certainly means underestimated 5f
moments (which is indeed what we observe in
Sec.~\ref{sec:moments}). We fix this deficiency by introducing an
empirical exchange field~$\boldsymbol{\Delta}_{fd}$ analogously to
the earlier computational studies of rare-earth systems
\cite{peters2014,shick2018}. The magnitude of this field is
estimated as $\Delta_{fd}\approx I_{fd} m_d$ \cite{peters2014}, where
$m_d$ is the magnetic moment due to the 6d electrons and $I_{fd}$ is
intra-atomic exchange integral. The magnetic moment is
approximated by its LSDA value, $m_d\approx 0.24$~\muB\ (see
Table~\ref{loc_ham} for the spin-resolved filling of the 6d bands),
the exchange integral is estimated by atomic calculations,
$I_{fd}\approx 0.15$~eV/\muB\ \cite{brooks1983}. This yields $I_{fd}
m_d\approx 36$~meV and we explore the LDA+DMFT solutions for
$\Delta_{fd}$ varied around this value.

The absence of $\boldsymbol{\Delta}_X$ is a disadvantage of the
spin-restricted parent band structure. Its advantage, on the other hand,
is that the double-counting correction in Eq.~\eqref{coulumbU}
reduces to a single number, $U_H$, since the spin-dependent part,
$U_X$, vanishes. One possible approximation to the double counting is
the so-called fully localized limit (FLL),
\begin{equation}
\label{eq:UH_FLL}
U_{H}^\text{FLL} =
U(n_{f}- 1/2) - J(n_{f}-1)/2\,,
\end{equation}
where $n_{f}$ is the self-consistently determined number of 5f
electrons \cite{solovyev1994,anisimov1997b}. In our calculations, it
turned out  that this $U_{H}^\text{FLL}$ severely overestimates the
number of 5f electrons, resulting in $n_{f}\approx4$. We hence employ an
alternative strategy: we choose $U_{H}$ such that the number of 5f
electrons remains close to its LDA value ($n_f=2.72$,
Table~\ref{loc_ham}) also in the LDA+DMFT solution to simulate charge
self-consistency \cite{amadon2008,havela2018}. This condition implies
$U_H\approx 3$~eV. We note in passing that the FLL formula,
Eq.~\eqref{eq:UH_FLL}, gives 3.93~eV for $n_f=2.72$, 4.41~eV for
$n_f=3$, and 2.71~eV for $n_f=2$.


\subsection{LSDA+DMFT}
\label{sec:methodLSDADMFT}

As discussed above, using spin-restricted LDA as the parent
band structure has two deficiencies: underestimated spin polarization
of the 6d (and other) bands, and missing exchange field due to 6d
moments acting on the 5f electrons. We dealt with the second issue
empirically, but we did not address the first one yet. We attempt to
do so by using the spin-polarized (LSDA) solution as the parent band
structure. This way, all non-5f bands are potentially spin-polarized,
which enhances the polarization of the bath and of the bath--5f
hybridization in the auxiliary impurity model, Eq.~\eqref{nonintsiam},
when compared to LDA+DMFT described in Sec.~\ref{sec:methodLDADMFT}.

Although it may seem that the LSDA parent band structure also provides
an improved estimate of the local exchange
field~$\boldsymbol{\Delta}_X$, it is not so, since the LSDA exchange
field combines the 6d--5f exchange (tens of meV) with the 5f--5f
exchange (about 1~eV). The latter has to be removed by the
double-counting correction $U_X$, which we know only
approximately. The FLL ansatz for the double counting $U_X$ reads as
\cite{anisimov1997b}
\begin{equation}
\label{eq:UX_FLL}
U_X^\text{FLL}
 = E^{\dn}_\text{FLL} - E^{\up}_\text{FLL}
 = J (n^{\up}_f - n^{\dn}_f)\,,
\end{equation}
where
\begin{equation}
E^{\sigma}_\text{FLL}=U(n_{f}- 1/2) -J(n_{f}^\sigma-1/2)\,,
\end{equation}
which we find to overcorrect the LSDA 5f--5f exchange. With the LSDA
occupation numbers (Table~\ref{loc_ham}) and with $J=0.59$~eV,
the double counting $U_X^\text{FLL}$ becomes 1.19~eV whereas the LSDA
exchange is only $\Delta_X=0.98$~eV (Table~\ref{loc_ham}).

Instead of using Eq.~\eqref{eq:UX_FLL} or any other similar formula,
we again employ the approach introduced in 
Sec.~\ref{sec:methodLDADMFT}, that is, we select $U_X$ such that
$\Delta_{fd}=\Delta_X-U_X\approx I_{fd} m_d \approx 36$~meV. Since
$\Delta_X$ is a parameter of the local Hamiltonian, it remains
constant during the DMFT self-consistency iterations as follows from
the derivation presented in Appendix~\ref{app:Hloc}, hence
$\Delta_{fd}$ and $U_X$ remain constant as well.

For the spin-independent part of the double-counting correction, we choose
$U_H = 3.3$~eV. This value is $0.3$~eV larger than in LDA+DMFT
because the average position of the 5f level is approximately $0.3$~eV
higher in the ferromagnetic LSDA solution compared to the non-magnetic
LDA solution (Table~\ref{loc_ham}).


\section{Results}
\label{sec:results}

\subsection{Magnetic moments}
\label{sec:moments}

\begin{figure}
\hbox to\linewidth{\hfill%
\includegraphics[width=\figfrac\linewidth]{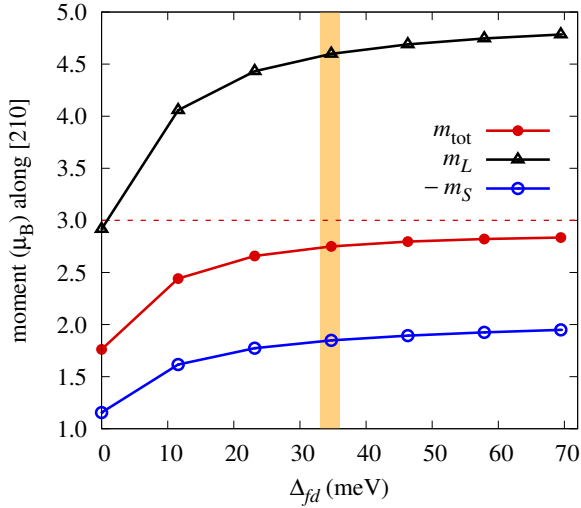}%
\hfill}
\caption{\label{fig:momLDADMFT210xc}The total magnetic moment (red),
  the orbital (black) and spin (blue) contributions to the magnetic
  moment of the 5f shell as functions of the exchange field $\Delta_{fd}$
  applied along the [210] direction in the LDA+DMFT calculations. The
  realistic value of $\Delta_{fd}$ is marked by the orange stripe, the
  experimental magnetic moment is indicated by the dashed line
  \cite{kolomiets2015}.}
\end{figure}

The method outlined in the preceding sections is not entirely
self-contained -- there are several semi-empirical parameters, 
such as the Coulomb parameters $F_k$, the double-counting
correction $U_H$, and the exchange field $\Delta_{fd}$. 
Especially the exchange field was estimated only roughly 
and hence we decided to explore a range of values around 
this estimate.


In Figure~\ref{fig:momLDADMFT210xc} we show how the magnetic moments 
depend on $\Delta_{fd}$ in the LDA+DMFT calculations when $\Delta_{fd}$
is applied in plane, along the $[210]$ direction, which corresponds to
the [210] ferromagnetic state
\footnote{We choose the [210] in-plane direction of magnetization instead of
  the experimental [100] direction due to technical limitation of our
  impurity solver. We have checked that this change does not
  significantly affect the ordered magnetic moments or the spectra in
  LSDA (Table~\ref{tab:mom_dft}).}%
. The orbital and spin contributions to the magnetic moment are
antiparallel as expected for 5f filling smaller than~7. At
$\Delta_{fd}=0$~meV, the total magnetic moment is clearly
underestimated (1.76~\muB), which confirms our earlier reasoning that
some exchange field has to be introduced. As the exchange field
increases, the magnetic moment quickly increases too, it reaches
2.75~\muB\ at $\Delta_{fd}=35$~meV, at which point it is already very
close to the saturation value $\approx 2.88$~\muB. The quick
saturation of the moments is a convenient feature -- an inaccuracy in
estimating the realistic value of $\Delta_{fd}$ translates to only a
minor uncertainty of the computed magnetic moments. The moments and 5f
filling at the realistic value of $\Delta_{fd}$ are compared to the
LSDA solution and to experiments in Table~\ref{tab:mom_dmft}.


\begin{table}[b]
\caption{\label{tab:mom_dmft}The orbital and spin magnetic moments in
  uranium 5f shells, $m_S$ and $m_L$ (in \muB), the total
  magnetic moment in the unit cell $m_\text{tot}$ (in \muB), the
  occupation of the 5f shells $n_f$, and the Sommerfeld
  coefficient $\gamma$ (in mJ/mol$\cdot$K${}^2$). The moments and the
  5f filling correspond to the maximally localized Wannier
  functions. The experimental $m_\text{tot}$ is taken from
  \cite{kolomiets2015}, the experimental $\gamma$ from \cite{honma2000}.}
\begin{ruledtabular}
\begin{tabular}{lcccccccc}
 & $U_H$ & $\Delta_{fd}$ & dir.&
 $m_{S}$ & $m_{L}$ & $ m_\text{tot}$ & 
 $n_f$ & $\gamma$ \\
\hline
LSDA      & -- & --  & [210] &$-1.96$& $2.79$ & $0.65$ & 2.72&24.5\\
LSDA      & -- & --  & [001] &$-2.00$& $2.89$ & $0.70$ & 2.72&21.2\\[1ex]
%
LDA+DMFT  & 3.0 & 35 & [210] &$-1.85$& $4.60$ & $2.75$ & 2.76&\08.2\\
LDA+DMFT  & 3.0 & 35 & [001] &$-1.91$& $4.87$ & $2.96$ & 2.74&\07.7\\[1ex]
LSDA+DMFT & 3.3 & 35 & [210] &$-1.66$& $4.15$ & $2.30$ & 2.82&\07.2\\ 
LSDA+DMFT & 3.3 & 35 & [001] &$-1.59$& $3.89$ & $2.12$ & 2.80&\07.5\\[1ex]
experiment   & & &     &        &        & 3.07 &     & 11.0\\
\end{tabular}
\end{ruledtabular}
\end{table}

\begin{figure}
\hbox to\linewidth{\hfill%
\includegraphics[width=\figfrac\linewidth]{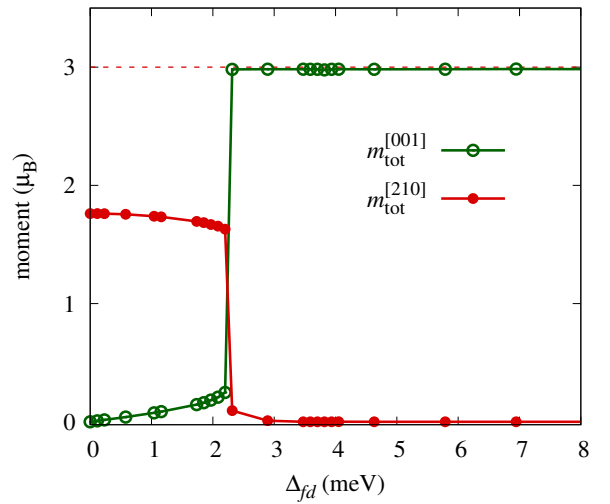}%
\hfill}
\caption{\label{fig:momLDADMFT001xc}Projection of the total magnetic
  moment to the [210] direction (red) and to the [001] direction
  (green) when the exchange field $\Delta_{fd}$ is applied along the
  [001] direction and the LDA+DMFT calculations are started from the
  LDA solution with $\Sigma(z)=0$. The orbital and spin moments (not
  shown) behave similarly as in Fig.~\ref{fig:momLDADMFT210xc}.}
\end{figure}

Analogous calculations were performed also for the exchange field
$\Delta_{fd}$ applied along the out-of-plane [001] direction. In this
case, the ferromagnetic state parallel to the exchange field is stable
only above some critical value of $\Delta_{fd}$, see
Fig.~\ref{fig:momLDADMFT001xc}. Above this value, the magnetic moment
very quickly saturates, much faster that in
Fig.~\ref{fig:momLDADMFT210xc}. For smaller values of $\Delta_{fd}$,
the DMFT iterations converge to a nearly in-plane state with just a
small out-of-plane tilt of the magnetic moments. For a range of
$\Delta_{fd}$ values we get two stationary solutions, one nearly
in-plane and the other out-of-plane, depending on the starting point
of the DMFT iterations. Figure~\ref{fig:momLDADMFT001xc} shows
calculations that were started at a given $\Delta_{fd}$ from the LDA
state with $\Sigma(z)=0$. The transition from the in-plane to
out-of-plane state then occurs at $\Delta_{fd}\approx
2.2$~meV. Calculations starting from the [001] ferromagnetic state
converge to the out-of-plane state already at $\Delta_{fd}\gtrsim
0.6$~meV (not shown).

Unfortunately, we cannot determine which of the two stationary states
found for $\Delta_{fd}$ between 0.6~meV and 2.2~meV 
is the ground state because we cannot reliably 
evaluate the total energy in our LDA+DMFT implementation. For the same
reason, we cannot estimate the magnetocrystalline anisotropy
energy. We can, however, conclude that the response of the magnetic
moments to $\Delta_{fd}$ as observed in LDA+DMFT is consistent with
the experimental finding that the easy axis is oriented in
plane. Starting from the paramagnetic state ($\Delta_{fd}=0$) and
cooling down, the system always ends up in the in-plane state, since
the moments exhibit an instability toward in-plane direction.
Increasing in-plane moment increases in-plane $\Delta_{fd}$, which
stabilizes the in-plane state further.


\begin{figure}
\hbox to\linewidth{\hfill%
\includegraphics[width=\figfrac\linewidth]{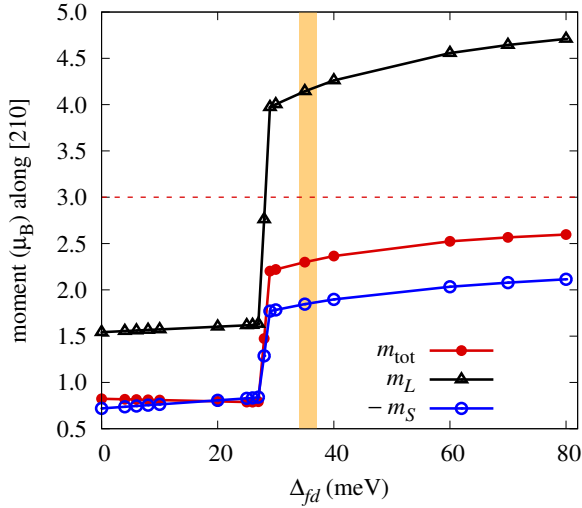}%
\hfill}
\caption{\label{fig:momLSDADMFT210xc}Variation of magnetic moments
  with $\Delta_{fd}$, computed using the LSDA+DMFT method when the parent
  band structure is polarized in the [210] direction. Compare with
  Fig.~\ref{fig:momLDADMFT210xc}.}
\end{figure}

\begin{figure}
\hbox to\linewidth{\hfill%
\includegraphics[width=\figfrac\linewidth]{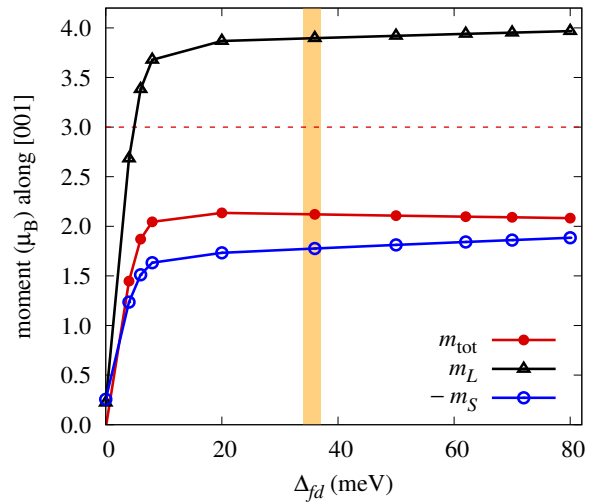}%
\hfill}
\caption{\label{fig:momLSDADMFT001xc}The same plot as in
  Fig.~\ref{fig:momLSDADMFT210xc}, only the parent LSDA band structure is
  polarized in the [001] direction. To be compared with
  Fig.~\ref{fig:momLDADMFT001xc}.}
\end{figure}

The magnetic moments computed using LSDA+DMFT, with the spin-dependent
part of the double-counting correction $U_X$ varied to reproduce the
same range of $\Delta_{fd}$ as explored above, are presented in
Figs.~\ref{fig:momLSDADMFT210xc} and~\ref{fig:momLSDADMFT001xc} for
the in-plane and out-of-plane orientation of the
LSDA polarization. As in the LDA+DMFT, the total magnetic moments
relatively quickly saturate with increasing $\Delta_{fd}$, and the
saturation is again faster in the [001] state than in the [210] state.
Surprisingly, the saturated values of the total moments are noticeably
smaller than in the corresponding LDA+DMFT
calculations, by 15\% in the case of the [210] ferromagnet and by 30\% in
the case of the [001] ferromagnet (compare with
Figs.~\ref{fig:momLDADMFT210xc} and~\ref{fig:momLDADMFT001xc}). We
expected the opposite, since the LSDA parent band structure is
certainly more polarized than the LDA parent band structure -- besides
$\Delta_{fd}$ that is the same in both approaches by construction, the
LSDA has all non-5f bands spin split, which results in an enhanced
polarization of the hybridization function. Intuitively, this should
have induced a larger polarization in the 5f shell but the
calculations show that it does not.

The difference in the computed moments could in principle be due to a
difference in fillings of the 5f shell between the LDA+DMFT and
LSDA+DMFT solutions, but this is not the case either. The 5f~filling in
both methods is very close as can be checked in
Table~\ref{tab:mom_dmft} where we summarize our results for the
realistic setting of the exchange field $\Delta_{fd}$. We speculate
that the inaccurate LSDA+DMFT moments come from some artifact of the
static LSDA approximation, {\change possibly from an artificially broken
symmetry}. One suspect feature is the strong spin
dependence of the crystal-field parameters $B_{kq}$ in the local
Hamiltonian shown in Appendix~\ref{app:Hloc}. Another feature, for
which we do not have a clear explanation and which is likely to be
connected to the LSDA solution as well, is the jump in magnetic
moments near $\Delta_{fd}=30$~meV in Fig.~\ref{fig:momLSDADMFT210xc}.

{\change

\begin{figure}
\hbox to\linewidth{\hfill%
\includegraphics[width=\figfrac\linewidth]{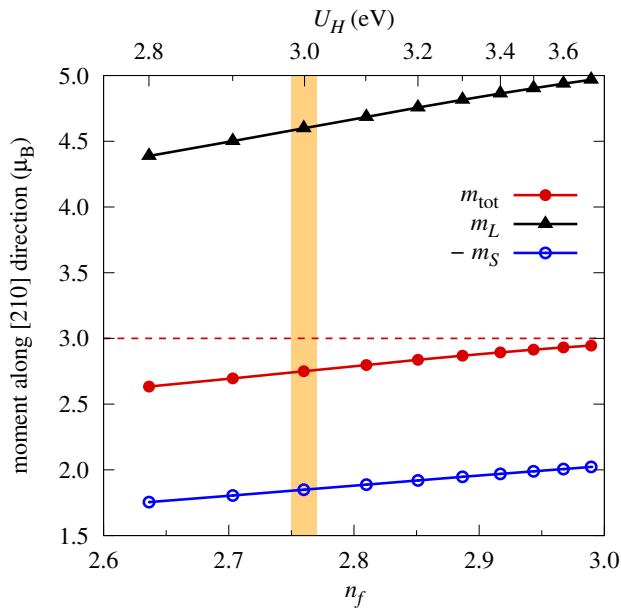}%
\hfill}
\caption{\label{fig:momLDADMFT210DC}\change The computed LDA+DMFT magnetic moments
  plotted as functions of the 5f filling $n_f$ that is
  varied by changing the double-counting correction $U_H$. The exchange field
  $\Delta_{fd}=35$~meV was applied along the [210] direction. The
  value $U_H=3.0$~eV employed throughout the paper is marked by the
  orange stripe.}
\end{figure}

Figures~\ref{fig:momLDADMFT210xc}--\ref{fig:momLSDADMFT001xc} show the
computed magnetic moments as functions of the exchange field 
$\Delta_{fd}$ for a fixed spin-independent part of the
double-counting correction~$U_H$. Although the employed values of
$U_H$ are well justified in Secs.~\ref{sec:methodLDADMFT}
and~\ref{sec:methodLSDADMFT}, it is useful to analyze the
sensitivity of the magnetic moments to changes of $U_H$ or,
equivalently, to changes of the 5f filling $n_f$. This sensitivity is
illustrated in Fig.~\ref{fig:momLDADMFT210DC} for the [210]
ferromagnetic state calculated with the LDA+DMFT method. The 
[001] ferromagnetic state and the results of the LSDA+DMFT method behave
analogously. The magnetic moments increase toward the experimentally
determined value with increasing $n_f$ but this route to improved
agreement with experiments does not have a solid physical
backing. Moreover, it would come at the cost of worsened 
agreement with the spectroscopic measurements, since increased $U_H$
would push the uranium 5f states to too large binding energies.

}


\subsection{Valence-band spectroscopy}
\label{sec:PES}

\begin{figure*}
\hbox to\linewidth{\hfill%
\includegraphics[width=0.90\linewidth]{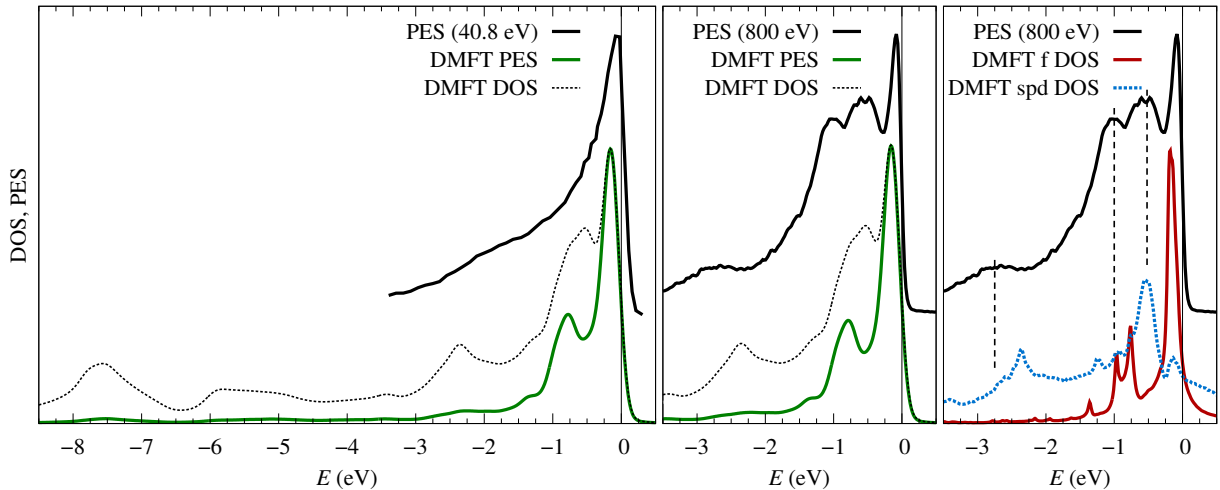}%
\hfill}
\caption{\label{fig:DOSPES} Experimental photoelectron spectra (black
  line) from \cite{gouder2001b} (left panel) and from
  \cite{fujimori2019,fujimori2020private} (middle panel) are compared
  to the LDA+DMFT estimate of the spectra (green line). A Gaussian
  broadening (FWHM 0.2~eV) is added to simulate the instrument resolution. The
  LDA+DMFT total DOS, subject to the same broadening, is shown for
  comparison ({\change dotted} line). In the right panel, we plot the
  orbital-resolved DOS without broadening (5f in red, sum of all
  others in {\change dotted} blue). All theoretical lines correspond to the [210]
  ferromagnet ($\Delta_{fd}=35$~meV).}
\end{figure*}

Two measurements of valence-band photoemission spectra of
UGa\textsubscript{2} can be found in the literature, the ultraviolet
photoemission spectrum \cite{gouder2001b} (UPS, shown in the left
panel of Fig.~\ref{fig:DOSPES}) and the soft-x-ray photoemission
spectrum \cite{fujimori2019} (SX-PES, shown in the middle panel of
Fig.~\ref{fig:DOSPES}). The UPS was measured on sputter-deposited
films at room temperature, that is, in the paramagnetic phase. The
maximum intensity was observed just below the Fermi level with a long
tail extending toward higher binding energies. The SX-PES was measured
on a freshly cleaved single crystal at $T=20$~K, that is, well below
the Curie temperature. The spectrum shows a narrow peak slightly below
the Fermi level accompanied with two broader features at $-0.5$~eV and
$-1.0$~eV, and an even broader hump can be discerned at $-2.8$~eV.

The two spectra are clearly different and the difference cannot be
ascribed to the lower resolution of the UPS spectra. The magnetic
order is also unlikely to cause such large changes, we certainly
do not see any evidence of that in DFT+DMFT calculations
(not shown), and the experiment does not detect any changes either
\cite{fujimori2019}. The more probable source of the differences is
the probing depth of the two experiments. The UPS used incident photons
with energy 40.8~eV (He II line), SX-PES used 800~eV (synchrotron
radiation), and hence the photoelectrons are emitted from deeper
layers in the bulk of the sample in the SX-PES measurements.

Since our calculations do not include any surface effects, they should
be closer to the SX-PES data. In Figure~\ref{fig:DOSPES} we show our
theoretically estimated photoelectron spectra at the appropriate
photon energies, calculated for the [210] ferromagnetic phase with
the LDA+DMFT method ($\Delta_{fd}=35$~meV, but the spectra are not
sensitive to variations of the 6d--5f exchange field). The spectra are
constructed as linear combinations of the \emph{orbital-resolved}
densities of states (DOS) weighted with photoionization cross sections
listed in \cite{yeh1985atomic}. According to these cross sections, the
5f DOS has by far the largest weight for both 40.8~eV and 800~eV
photon energies, and hence these photoemission measurements probe
mainly the 5f states.

The computed spectra display a main peak at $-0.15$~eV and
a satellite at $-0.8$~eV. The satellite has a considerably smaller
intensity than the features seen in the SX-PES and as such the theory
appears to be closer to the UPS spectra. The $-0.5$~eV and $-2.8$~eV features
observed in SX-PES do not show up in the theoretical PES, but there are
distinct peaks appearing at nearby energies in the LDA+DMFT
\emph{total} DOS (Fig.~\ref{fig:DOSPES}). They originate from
orbitals that have small photoionization cross sections. These
peaks are due to hybridized U 6d and Ga 4p bands at $-0.5$~eV, and
mainly Ga 4p bands at $-2.4$~eV. The distinct feature outside the
range probed by photoemission, at $-7.6$~eV, is due to Ga 4s. The fact
that SX-PES sees a signal where the theory places Ga 4p bands may be
an indication that the theory underestimates the hybridization between
Ga~4p and U~5f states. If the hybridization was stronger, some U 5f
DOS would possibly appear at the position of the Ga 4p states, but
that is just a speculation at this point.

\begin{figure}
\hbox to\linewidth{\hfill%
\includegraphics[width=\figfrac\linewidth]{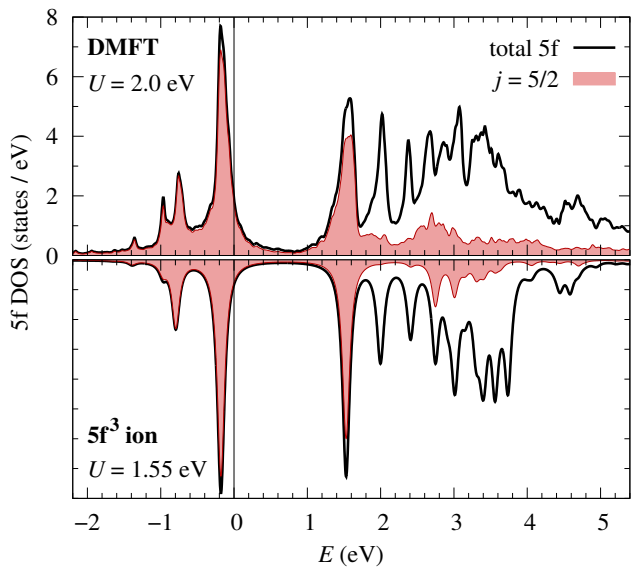}%
\hfill}
\caption{\label{fig:5fDOS}The uranium 5f DOS in the
  [210] ferromagnet from the LDA+DMFT method ($\Delta_{fd}=35$~meV)
  in the top panel is compared to the DOS from an atomic
  calculation (5f\textsuperscript{3} state) in the bottom panel (black
  lines). The parameter $F_0=U$ was reduced in the atomic calculation
  to mimic the screening effects incorporated in the LDA+DMFT
  method. The $j=5/2$ components of the 5f DOS are shown in red.}
\end{figure}

Photoemission experiments access only the occupied part of the
spectrum. The unoccupied part could be probed by inverse photoemission
(we are not aware of any such experiment being performed to date) or
by x-ray absorption spectroscopy (we discuss recent x-ray absorption
measurements at the uranium M\textsubscript{4,5} edges in
UGa\textsubscript{2} elsewhere \cite{kolomiets2021}). In
Figure~\ref{fig:5fDOS}, we analyze the complete (occupied and
unoccupied) 5f DOS from a theoretical perspective. We compare the
LDA+DMFT result with the DOS computed for a spherically symmetric
5f\textsuperscript{3} ion, {\change since three is the closest integer
value to the computed average 5f 
filling $n_f$ (Table~\ref{tab:mom_dmft}) and the probability of
finding the 5f shell in the 5f\textsuperscript{3} configuration
predicted by LDA+DMFT is large, namely 83\%. See Appendix~\ref{app:Mconv},
Eq.~\eqref{eq:fluctNf}, for the meaning of the fluctuating number
of 5f electrons.} We can achieve a very close correspondence of the
ionic and LDA+DMFT
densities of states when the Coulomb~$U$ in the ionic model
is reduced to $1.55$~eV compared to $2.0$~eV in LDA+DMFT. The higher
Slater parameters $F_k$ and the spin-orbit parameter $\zeta$ are
identical. This observation indicates that the 5f states in the
LDA+DMFT are very close to being fully localized, only their Coulomb
repulsion is screened more than it would be in the fully localized
Hubbard-I approximation. In addition, Fig.~\ref{fig:5fDOS} also shows
the $j=5/2$ component of the 5f~DOS to be compared with the shape of
the M\textsubscript{4} absorption line \cite{kolomiets2021}.

\begin{figure}
\hbox to\linewidth{\hfill%
\includegraphics[width=\figfrac\linewidth]{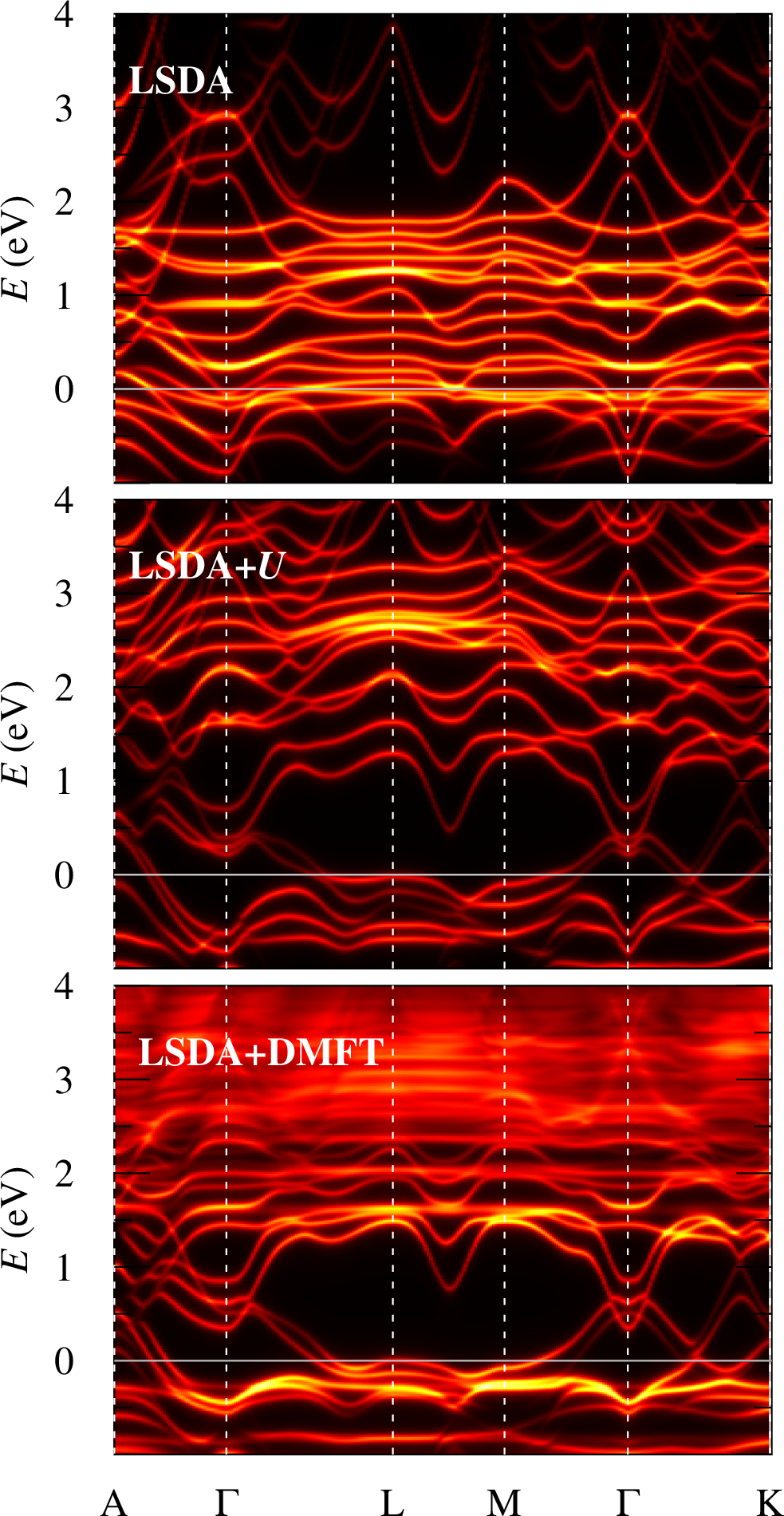}%
\hfill}
\caption{\label{fig:band}Momentum-resolved 5f spectral density. The
  electronic correlations are described with increasing 
  level of sophistication from top to bottom: LSDA, LSDA+$U$
  ($U=2.0$~eV, $J=0.59$~eV and the FLL double counting), and LSDA+DMFT
  (the same interaction parameters, and $U_{H}=3.3$~eV and
  $\Delta_{fd}=35$~meV). The [210] ferromagnetic state is shown in all
  three panels. The same $U$ and $J$ produce a larger gap between the
  occupied and unoccupied 5f states in LSDA+$U$ than in LSDA+DMFT.}
\end{figure}

Finally, in Fig.~\ref{fig:band} we present the momentum-resolved 5f
spectral density along high-symmetry directions in the Brillouin zone. We
compare different models for the electronic correlations, namely LSDA,
LSDA+$U$ and LSDA+DMFT, in the ferromagnetic state with magnetic
moments pointing along the [210] direction. The [001] ferromagnetic state
differs only in minor details. When the Hubbard term is included
(LSDA+$U$ and LSDA+DMFT), a gap between the occupied and unoccupied 5f
bands appears and the occupied 5f states move slightly away from the
Fermi level. Given the same interaction parameters ($U$ and $J$, or
$F_k$), this gap is larger in LSDA+$U$, which indicates that the
screening of the Coulomb parameters is stronger in LSDA+DMFT than in
LSDA+$U$. The situation is analogous to Fig.~\ref{fig:5fDOS} since the
$U$-induced potential in LSDA+$U$ has the form of a \emph{ionic}
Hartree--Fock approximation. Another difference between the LSDA+DMFT and
LSDA+$U$ electronic structure is the incoherent character of the 5f
states visible in the LSDA+DMFT solution, starting approximately $2.5$~eV
above the Fermi level.


\subsection{Sommerfeld coefficient}
\label{sec:sommerfeld}

Figure~\ref{fig:band} illustrates that the Fermi level cuts right
through the 5f bands in LSDA, which is accompanied by a high density
of states at the Fermi level and, subsequently, by a large Sommerfeld
coefficient of the electronic specific heat $\gamma$. Indeed, LSDA
predicts $\gamma>20$~mJ/mol$\cdot$K\textsuperscript{2}
(Table~\ref{tab:mom_dmft}), which is at odds with the experimental value
11~mJ/mol$\cdot$K\textsuperscript{2} \cite{honma2000}. In DFT+DMFT
(and in LSDA+$U$ as well), the 5f states move away from the Fermi
level toward higher binding energies and the coefficient $\gamma$ is
reduced to approximately 8~mJ/mol$\cdot$K\textsuperscript{2}
(Table~\ref{tab:mom_dmft}), yielding 
a considerably better agreement with experiments. The computed
Sommerfeld coefficient should be smaller than observed in experiments
since we do not take into account any enhancement due to phonons. We
do not observe much variation of $\gamma$ when changing the
orientation of the magnetic moments or when alternating the parent
band structure (Table~\ref{tab:mom_dmft}).


\section{Conclusions}
\label{sec:summary}

We have studied the electronic structure and magnetic properties of
the ferromagnetic compound UGa\textsubscript{2} using the DFT+DMFT
method, and compared our results with more approximate
electronic-structure methods. We have found that our implementation of
the DFT+DMFT method reproduces the experimentally observed large
magnetic moments as well as the sign of the magnetocrystalline
anisotropy energy, when the exchange interaction between uranium 6d
and 5f states is included in a semi-empirical manner. This is done
either in the form of an extra potential acting on the 5f states or in
the form of a spin-polarized double-counting correction. We have
compared two formulations of the DFT+DMFT method, one keeping the
non-5f states spin restricted (LDA), and the other allowing their spin
polarization (LSDA). Of the two, the LDA-based variant was found to
provide more consistent results. It is a future work to investigate
how the semi-empirical approach to the 6d--5f exchange could be
improved toward a fully first-principles method.

Besides the magnetic properties, we have also modeled the valence-band
photoemission spectra on the basis of the DFT+DMFT density of states. We were
not able to fully explain the differences between the two published
photoemission experiments \cite{gouder2001b,fujimori2019} but we could
understand how the electron-electron correlations move the 5f states
slightly away from the Fermi level, which is in accord with both photoemission
spectra as well as with the observed small Sommerfeld coefficient of
the electronic specific heat. With the aid of the DFT+DMFT method, it
is thus possible to reconcile large magnetic moments and a small
Sommerfeld coefficient with the 5f spectral density in the close
vicinity of the Fermi level.

Our calculations indicate a close-to-localized uranium 5f states in
UGa\textsubscript{2}. From the comparison to the experimental
photoemission spectra we deduce that the tendency to localization is
probably slightly overestimated in our theoretical description. Such a
tendency is to be expected for the employed impurity solver that
implements a form of expansion around the atomic limit.


\begin{acknowledgments}
The work was supported by the Czech Science Foundation under
the grants No.~18-02344S and No.~21-09766S. We thank S.-i. Fujimori for
experimental data, and L.~Havela, J.~Kune\v s and A.~B. Shick for fruitful
discussions. Computational resources were partially supplied by the
project ``e-Infrastruktura CZ'' (e-INFRA LM2018140) provided within
the program Projects of Large Research, Development and Innovations
Infrastructures.
\end{acknowledgments}


\appendix


\section{Parameters of DFT calculations}
\label{app:DFT_params}

\begin{table}[b]
\caption{\label{tab:mom_dft}The orbital and spin magnetic moments in
  uranium 5f shell, $m_S$ and $m_L$ (in \muB), the total
  magnetic moment in the unit cell $m_\text{tot}$ (in \muB), the
  occupation of the 5f shell $n_f$, and the Sommerfeld
  coefficient $\gamma$ (in mJ/mol$\cdot$K\textsuperscript{2}). The 5f
  magnetic moments and the 5f filling correspond to the atomic
  (muffin-tin) spheres.}
\begin{ruledtabular}
\begin{tabular}{lcccccc}
 & direction&
 $m_{S}$ & $m_{L}$ & $ m_\text{tot}$ & $n_f$ & $\gamma$ \\
\hline
LDA       &  --   &  --   &   --   &   --   & 2.45& 43.9\\[1ex]
LSDA      & [100] &$-1.82$& $2.67$ & $0.57$ & 2.51& 24.7\\
LSDA      & [210] &$-1.82$& $2.64$ & $0.54$ & 2.50& 26.7\\
LSDA      & [001] &$-1.86$& $2.72$ & $0.57$ & 2.50& 22.5\\
\end{tabular}
\end{ruledtabular}
\end{table}

To perform all DFT calculations presented in this paper, we employed the
WIEN2k package \cite{wien2k} that implements linearized augmented plane-wave
method and its extensions. It combines a scalar-relativistic
description with spin-orbit coupling \cite{koelling1977}. All calculations were performed
at the experimental lattice constants $a=4.213$~\AA\ and 
$c=4.020$~\AA, reported in \cite{andreev1978}, with the following
parameters: the radii of the muffin-tin spheres were 
$R_{\rm MT}({\rm U})=2.80\, a_{\rm B}$ for uranium atoms and $R_{\rm
  MT}({\rm Ga})=2.25\, a_{\rm B}$ for gallium atoms, the Brillouin
zone was sampled with 6137 k points (900 k points in the irreducible
wedge), and the basis-set cutoff $K_{\rm max}$ was defined with $R_{\rm
  MT}({\rm Ga})\times K_{\rm max}=10.0$. The default basis set with
local orbitals for semicore states (U 6s, 6p, and Ga 3d) was used
in all cases.

In Table~\ref{tab:mom_dft}, we list the orbital and spin magnetic
moments of the uranium 5f shell, the total magnetic moment of the unit
cell, the filling of the 5f shell, and the Sommerfeld coefficient for
three ferromagnetic states with moments pointing along different
crystallographic axes. The moments and the filling of the 5f shell
correspond to the muffin-tin sphere, they can be compared to the
values computed for the maximally localized Wannier functions shown in
Table~\ref{tab:mom_dmft}. The largest components of the total moment
quoted in Table~\ref{tab:mom_dft} are the 5f moments, a sizable
contributions come also from the spin moments in the U~6d states
($\approx-0.1$~\muB) and in the interstitial ($\approx-0.2$~\muB). The
moments induced at Ga atoms are negligible.

The maximally localized Wannier functions for the DMFT calculations
were found with the Wannier90 code \cite{mostofi2008}. The spread
minimization was performed on $16\times16\times16$ mesh of k
points. Since there are no gaps in the spectrum above the Fermi level,
disentanglement was necessary \cite{souza2001}. We used 62 Bloch
states on input, which corresponds to the energy window from $-10$~eV
to 24~eV. (Our largest tight-binding models, that is, those actually
used for the DMFT calculations, have 48 Wannier functions). The frozen
inner window extended to 6~eV (3~eV for the smallest model listed
in Table~\ref{loc_ham}), going higher meant that the centers
of the Wannier functions started drifting away from the atomic
centers, which is undesirable in our application that assumes the
Wannier functions to be atomic-like. In the model used for the DMFT
calculations, the original WIEN2k bands were represented
perfectly up to 6~eV above the Fermi level, the match was still very
good up to approximately 12~eV, and above that the correspondence
quickly deteriorated.


\section{Construction of the impurity model}
\label{app:bath_discretization}

Here we discuss how the parameters of the finite
impurity model, Eq.~\eqref{nonintsiam}, are found so
that the model matches the effective medium (the bath) as closely
as possible. The impurity Hamiltonian has the form a block matrix
\begin{equation}
\label{eq:Himp_mtrx}
\mathbb{H}_\text{imp}=
\begin{pmatrix}
\mathbb{H}_\text{loc} & \mathbb{V}_1 & \mathbb{V}_2 & \mathbb{V}_3 & \cdots\\
\mathbb{V}_1^{\dag}   & \mathbb{H}_\text{bath}^{(1)} & 0 & 0 & \cdots\\
\mathbb{V}_2^{\dag}   & 0 & \mathbb{H}_\text{bath}^{(2)} & 0 & \cdots\\
\mathbb{V}_3^{\dag}   & 0 & 0 & \mathbb{H}_\text{bath}^{(3)} & \cdots\\
\vdots   & \vdots & \vdots & \vdots & \ddots
\end{pmatrix},
\end{equation}
where all blocks are $14\times 14$ square matrices. The local
Hamiltonian~$\mathbb{H}_\text{loc}$ contains a strong spin-orbit
coupling which does not commute with the hybridization function that
follows the crystal symmetry. Therefore, the problem cannot be
simplified to diagonal matrices.

If there is only one $\mathbb{H}_\text{bath}$ block, all three matrices
$\mathbb{H}_\text{loc}$, $\mathbb{H}_\text{bath}$ and $\mathbb{V}$ can
be determined by comparing the large $z$ asymptotics of the local
block of the impurity Green's function,
\begin{equation}
\label{eq:Gloc}
\mathbb{G}_\text{loc}(z)=\Bigl[z\mathbb{I} -\mathbb{H}_\text{loc}
 -\sum_i\mathbb{V}_i\bigl(z\mathbb{I} -\mathbb{H}_\text{bath}^{(i)}\bigr)^{-1}\mathbb{V}_i^{\dag}
\Bigr]^{-1},
\end{equation}
to the asymptotics of the bath Green's function defined as
\begin{equation}
\label{eq:bathGF}
\mathbb{G}(z)=\bigl[G_f^{-1}(z)+\Sigma(z)\bigr]^{-1}\,.
\end{equation}
Here $G_f(z)$ is the 5f block of the local Green's function
$G(z)$ from Eq.~\eqref{eq:localGFcrystal}. We refer the reader to
\cite{kolorenc2015b} for details. For larger
impurity models, like Eq.~\eqref{eq:Himp_mtrx}, this strategy leads to
an unsolvable set of polynomial equations for the $14\times 14$ square
matrices. To overcome the problem, we combine two shorter asymptotic
expansions, one for the Green's function as before, and one for the
hybridization function.

The asymptotic expansion of the local block of the impurity Green's
function $\mathbb{G}_{\rm loc}(z)$ starts as
\begin{equation}
\mathbb{G}_\text{loc}(z) = \frac{\mathbb{I}}z
+ \frac{\mathbb{H}_\text{loc}}{z^2}
+O(z^{-3})\,,
\label{eq:GFimp_asympt}
\end{equation}
and the analogous expansion of the hybridization function
\begin{equation}
\Delta_\text{imp}=z\mathbb{I}-\mathbb{H}_\text{loc}
 -\mathbb{G}_\text{loc}^{-1}(z)
\end{equation}
starts as
\begin{multline}
\Delta_\text{imp}=\sum_i\mathbb{V}_i\bigl(z\mathbb{I}
-\mathbb{H}_\text{bath}^{(i)}\bigr)^{-1}\mathbb{V}_i^{\dag}
=\\
\sum_i\biggl[\frac{\mathbb{V}_i \mathbb{V}_i^{\dag}}{z}
+\frac{\mathbb{V}_i\, \mathbb{H}_\text{bath}^{(i)}\mathbb{V}_i^{\dag}}{z^2}
\biggr]+O(z^{-3})\,.
\label{eq:hybrimp_asympt}
\end{multline}
From the other side, the bath Green's function,
Eq.~\eqref{eq:bathGF}, reads in the spectral representation as 
\begin{equation}
\mathbb{G}(z)
=\int\frac{\mathbb{A}(\epsilon)}{z-\epsilon}\,\rmd\epsilon\,,
\label{eq:GF0spectral}
\end{equation}
where we introduced the spectral density
\begin{equation}
\mathbb{A}(\epsilon)
=\frac{\mathbb{G}(\epsilon-\rmi0)-\mathbb{G}(\epsilon+\rmi0)}{2\pi\rmi}\,.
\end{equation}
The asymptotic expansion of the bath Green's function is obtained by
expanding the denominator in Eq.~\eqref{eq:GF0spectral},
\begin{equation}
\mathbb{G}(z)=\sum_{n=0}^{\infty}\frac{\mathbb{M}_{n}}{z^{n+1}}\,,
\quad
\mathbb{M}_{n}=\int\epsilon^{n} \mathbb{A}(\epsilon)\,\rmd\epsilon\,,
\label{eq:GF0_asympt}
\end{equation}
where $\mathbb{M}_n$ are moments of the spectral density. The
spectral density $\mathbb{A}(\epsilon)$ is a hermitian matrix and
hence its moments are hermitian matrices as well. We immediately see
that
\begin{equation}
\mathbb{H}_\text{loc}=\mathbb{M}_1\,.
\label{eq:Hloc_solution}
\end{equation}

The spectral representation of the hybridization function
corresponding to $\mathbb{G}(z)$, that is, of
$\Delta(z)=z\mathbb{I}-\mathbb{M}_1-\mathbb{G}^{-1}(z)$, can be written as
\begin{equation}
\Delta(z)
=\int\frac{\mathbb{B}(\epsilon)}{z-\epsilon}\,\rmd\epsilon\,,
\label{eq:hybrspectral}
\end{equation}
where the spectral density is defined as
\begin{equation}
\mathbb{B}(\epsilon)
=\frac{\Delta(\epsilon-\rmi0)-\Delta(\epsilon+\rmi0)}{2\pi\rmi}
\,.
\end{equation}
Now we split the support of $\mathbb{B}(\epsilon)$
to as many segments as many $\mathbb{H}_\text{bath}^{(i)}$ blocks we
wish (or can afford) to have,
\begin{equation}
\Delta(z)=\sum_i\Delta_i(z)\,,\text{ where }
\Delta_i(z)=\int_{\epsilon_i}^{\epsilon_{i+1}}
  \frac{\mathbb{B}(\epsilon)}{z-\epsilon}\,\rmd\epsilon
\end{equation}
with $\epsilon_i<\epsilon_{i+1}$, and we pair each $\Delta_i$ with one
summand in Eq.~\eqref{eq:hybrimp_asympt}. The splitting can be
arbitrary or it can be guided by an insight into the structure of the
hybridization function -- the individual
$\mathbb{H}_\text{bath}^{(i)}$ blocks can be aligned with groups of
bands. In UGa\textsubscript{2}, the hybridization
below the Fermi level comes mainly from Ga 4s and 4p bands, and in the
first $\approx 6$~eV above the Fermi level it is dominated by U 6d bands.

The asymptotic expansion at the individual intervals reads as
\begin{equation}
\Delta_i(z)=\sum_{n=0}^{\infty}\frac{\mathbb{N}_{n}^{(i)}}{z^{n+1}}\,,
\quad
\mathbb{N}_{n}^{(i)}=\int_{\epsilon_i}^{\epsilon_{i+1}}\epsilon^{n}
  \mathbb{B}(\epsilon)\,\rmd\epsilon\,.
\label{eq:hybr_asympt}
\end{equation}
Comparing Eqs.~\eqref{eq:hybrimp_asympt} and~\eqref{eq:hybr_asympt},
the blocks of $\mathbb{H}_\text{imp}$ can be written in terms of
the moments $\mathbb{N}_{n}^{(i)}$ as
\begin{subequations}
\label{eq:bath_solution}
\begin{align}
\mathbb{V}_i&=\mathbb{V}_i^{\dag}=\sqrt{\mathbb{N}_0^{(i)}}\,,\\
\mathbb{H}_\text{bath}^{(i)}&=\mathbb{V}^{-1}\mathbb{N}_1^{(i)}
\bigl(\mathbb{V}^{\dag}\bigr)^{-1}\,,
\end{align}
\end{subequations}
which, together with Eq.~\eqref{eq:Hloc_solution}, concludes the
construction of the impurity model $\mathbb{H}_\text{imp}$ from
the local Green's function $G(z)$. Optionally, we can diagonalize the blocks
$\mathbb{H}_\text{bath}^{(i)}$ to make their interpretation more
straightforward and to arrive at the form of the impurity model used
in Eq.~\eqref{nonintsiam}. The corresponding transformations are
\begin{equation}
\mathbb{H}_\text{bath}^{(i)}\to
  \mathbb{C}_i^{-1}\mathbb{H}_\text{bath}^{(i)}\mathbb{C}_i\,,\quad
\mathbb{V}_i\to \mathbb{V}_i\mathbb{C}_i\,,
\end{equation}
where $\mathbb{C}_i$ are the appropriate unitary matrices and the new
$\mathbb{V}_i$ are no longer hermitian. By construction, the
eigenvalues of $\mathbb{H}_\text{bath}^{(i)}$ are confined to
intervals $(\epsilon_i,\epsilon_{i+1})$.

\begin{figure}
\includegraphics[width=0.85\linewidth]{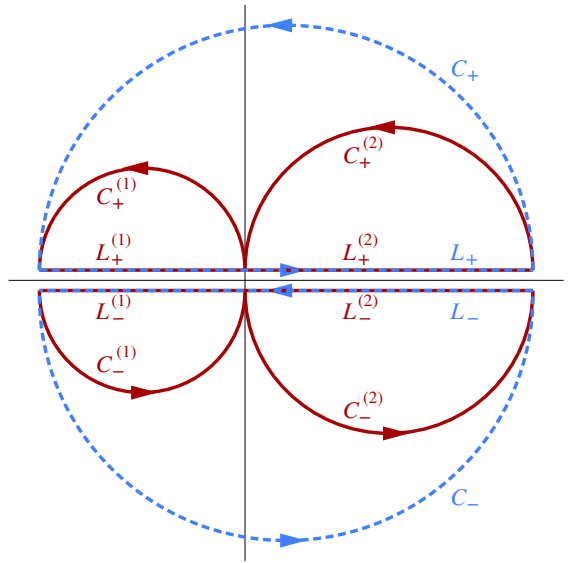}
\caption{\label{fig:moment_contours}Contours in the
  complex plane used for integration of the moments~$\mathbb{M}_1$
  ({\change dashed} blue) and $\mathbb{N}_n^{(i)}$ (red). Line
  segments are denoted as $L_{\pm}$, half circles as $C_{\pm}$.}
\end{figure}

For the purpose of their actual evaluation, the moments are expressed 
in terms of contour integrals in the complex plane. Using the path
segments sketched in Fig.~\ref{fig:moment_contours}, we have
\begin{align}
\mathbb{M}_1
={}&\frac1{2\pi\rmi}\biggl[\int_{-L_-}-\int_{L_+}\biggr]\,
  z\,\mathbb{G}(z)\, \rmd z\nonumber\\
&=\frac1{2\pi\rmi}\biggl[\int_{C_-}+\int_{C_+}\biggr]\,
  z\,\mathbb{G}(z)\, \rmd z\,,\\
%
\mathbb{N}_n^{(i)}
={}&\frac1{2\pi\rmi}\biggl[\int_{-L_-^{(i)}}-\int_{L_+^{(i)}}\biggr]\,
  z^n \bigl[z\mathbb{I}-\mathbb{M}_1-\mathbb{G}^{-1}(z)\bigr] \rmd z
\nonumber\\
&=\frac1{2\pi\rmi}\biggl[\int_{C_-^{(i)}}+\int_{C_+^{(i)}}\biggr]\,
  z^n \bigl[z\mathbb{I} -\mathbb{M}_1-\mathbb{G}^{-1}(z)\bigr] \rmd z\,,
\end{align}
where the integral over the ({\change dashed} blue) circle $C=C_-\cap
C_+$ encloses the 
entire support of $\mathbb{A}(\epsilon)$ and the integrals over the (red)
circles $C^{(i)}=C_-^{(i)}\cap C_+^{(i)}$ enclose the intervals
$(\epsilon_i,\epsilon_{i+1})$. During the DMFT calculations, the
self-energy is thus evaluated along the circles $C$ and $C^{(i)}$, and also
along one additional semicircle in the upper half plane to compute the
number of electrons in the primitive cell and to adjust the Fermi
level. An alternative to the circle $C$, which serves for evaluation
of $\mathbb{H}_\text{loc}=\mathbb{M}_1$, is described in
Appendix~\ref{app:Hloc}.

In the DFT+DMFT calculations of UGa\textsubscript{2} discussed in the
paper, we used three intervals $(\epsilon_i,\epsilon_{i+1})$, namely
$(-10, 0)$~eV, $(0,6)$~eV and $(6,12)$~eV. The hybridization above
12~eV was discarded, since our tight-binding Hamiltonians do not
accurately represent the original DFT bands that far above the Fermi
level (Appendix~\ref{app:DFT_params}).


\section{Asymptotics of the bath Green's function and the local Hamiltonian}
\label{app:Hloc}

At each k point, the tight-binding Hamiltonian $\hat H_{\mathbf k}$
can be divided into four blocks,
\begin{equation}
\hat H_{\mathbf k}=
\begin{pmatrix}
\hat H_{\mathbf k}^f & \hat T_{\mathbf k} \\
\hat T_{\mathbf k}^\dag       & \hat H_{\mathbf k}^{spd}
\end{pmatrix}\,,
\end{equation}
and the 5f block of the lattice Green's function can be written as
\begin{equation}
\hat G_{\mathbf k}^f(z)=\Bigl[
 z\hat{I}-\hat{H}_{\mathbf k}^f-\hat\Sigma(z)
 -\hat{T}_{\mathbf k}\bigl(
   z\hat{I} -\hat{H}^{spd}_{\mathbf k}\bigr)^{-1}
 \hat{T}_{\mathbf k}^{\dag}
\Bigr]^{-1}.
\end{equation}
Its asymptotic expansion reads as
\begin{equation}
\hat{G}_{\mathbf k}^f(z) = \frac{\hat{I}}z
+ \frac{\hat{H}_{\mathbf k}^f+\hat\Sigma(\infty)}{z^2}
+O(z^{-3})\,,
\end{equation}
where $\hat\Sigma(\infty)$ is the static part of the self-energy, which
is the leading term of the expansion
$\hat\Sigma(z)=\hat\Sigma(\infty)+O(z^{-1})$. For the bath Green's
function, Eq.~\eqref{eq:bathGF}, we need only the local element,
\begin{multline}
\hat G_f(z)=\frac1N\sum_{\mathbf k}\hat{G}_{\mathbf k}^f(z)\\
= \frac{\hat I}z 
+ \frac{N^{-1}\sum_{\mathbf k}\hat{H}_{\mathbf k}^f+\hat\Sigma(\infty)}{z^2}
+O(z^{-3})\,,
\end{multline}
respectively its inverse,
\begin{equation}
\hat G_f^{-1}(z)
=z\hat{I}-\frac1N\sum_{\mathbf k}\hat{H}_{\mathbf k}^f
-\hat\Sigma(\infty)+O(z^{-1})\,.
\end{equation}
Inserting this expression into the definition of the bath Green's
function, Eq.~\eqref{eq:bathGF}, yields
\begin{equation}
\mathbb{G}(z)=\frac{\hat{I}}z
 +\frac1{z^2}\frac1N\sum_{\mathbf k}\hat{H}_{\mathbf k}^f
+O(z^{-1})\,.
\end{equation}
The self-energy cancels out from the first moment of the corresponding
spectral density, and the moment thus equals to the local block of the
tight-binding Hamiltonian,
\begin{equation}
\mathbb{M}_1=\frac1N\sum_{\mathbf k}\hat{H}_{\mathbf k}^f=\mathbb{H}_\text{loc}\,,
\end{equation}
throughout the whole DMFT self-consistency loop.

\begin{table}[b]
\caption{\label{tab:spin_cef} Crystal-field
  parameters $B_{kq}^\sigma$, Eq.~\eqref{eq:Bkqs}, derived from the LSDA
  tight-binding Hamiltonian (s,p,d,f model). Spin-restricted
  parameters $B_{kq}$ computed from Eq.~\eqref{eq:Bkq} are the same as
  shown in Table~\ref{loc_ham}.}
\begin{ruledtabular}
\begin{tabular}{lcccc}
 & $B_{20}$&$B_{40}$&$B_{60}$&$B_{66}$\\
\hline
\multicolumn{5}{l}{ferromagnetic solution [001]}\\
restricted & $3.72$ &$\m0.0061$&  $-0.0043$& $-0.052$\\
spin $\up$ & $1.75$ & $-0.0024$& $\m0.0017$& $-0.107$\\
spin $\dn$ & $5.68$ &$\m0.0146$&  $-0.0100$&$\m0.003$\\[1ex]
\multicolumn{5}{l}{ferromagnetic solution [210]}\\
restricted & $3.92$ &$\m0.0262$& $-0.0011$& $-0.044$\\
spin $\up$ & $1.54$ &$\m0.0111$& $-0.0154$& $-0.009$\\
spin $\dn$ & $6.30$ &$\m0.0414$&$\m0.0133$& $-0.079$\\
\end{tabular}
\end{ruledtabular}
\end{table}

To extract the individual contributions to the Hamiltonian shown in
Eq.~\eqref{eq:HlocDecomp}, we can exploit the orthogonality of
operators $\hat I$, $\hat{\mathbf {l}} \cdot \hat{\mathbf {s}}$,
$\hat{\mathbf{s}}$ and $\hat O_{kq}$ as $14\times14$ matrices. We can
write
\begin{subequations}
\allowdisplaybreaks
\begin{align}
\epsilon_{f} 
  &=\tr(\mathbb{H}_\text{loc})/14\,,\\[.2ex]
\Delta_X^{\alpha} 
  &=\tr(\hat{s}_\alpha\,\mathbb{H}_\text{loc})/
     \tr(\hat{s}_\alpha\,\hat{s}_\alpha)\,,\quad\alpha=x,y,z\,,\\[.2ex]
\zeta
  &=\tr(\hat{\mathbf {l}}\cdot\hat{\mathbf {s}}\,\mathbb{H}_\text{loc})/
     \tr(\hat{\mathbf {l}}\cdot\hat{\mathbf {s}}\,
     \hat{\mathbf{l}}\cdot\hat{\mathbf {s}} )\,, \\[.2ex]
\label{eq:Bkq}
B_{kq}
  &=\tr(\hat O_{kq}\,\mathbb{H}_\text{loc})/
     \tr(\hat O_{kq}\hat O_{kq})\,.
\end{align}
In the case of spin-polarized electronic structure, spin-dependent
crystal-field parameters can be introduced as
\begin{equation}
\label{eq:Bkqs}
B_{kq}^\sigma
  =\tr(\hat O_{kq}\hat P_\sigma\,\mathbb{H}_\text{loc})/
     \tr(\hat O_{kq}\hat P_\sigma\hat O_{kq}\hat P_\sigma)\,,
\end{equation}
\end{subequations}
where $\hat P_\sigma$ is a projector to spin $\sigma$. Since the
operator $\hat O_{kq}$ is spin-independent, it commutes with $\hat
P_\sigma$ and we can simplify the denominator as
\begin{multline}
\tr(\hat O_{kq}\hat P_\sigma\hat O_{kq}\hat P_\sigma)=
\tr(\hat O_{kq}\hat O_{kq}\hat P_\sigma\hat P_\sigma)\\
=
\tr(\hat O_{kq}\hat O_{kq}\hat P_\sigma)=
\frac12\tr(\hat O_{kq}\hat O_{kq})\,.
\end{multline}
Consequently, the parameters $B_{kq}$ are averages of the
spin-dependent parameters $B_{kq}^\sigma$,
\begin{equation}
B_{kq}=\frac12\sum_\sigma B_{kq}^\sigma\,.
\end{equation}
The spin dependence of the crystal-field parameters derived from the
LSDA band structure is substantial, which is illustrated in
Table~\ref{tab:spin_cef}. Note that we do not attempt to remove the 5f
self-interaction from the crystal-field potential
\cite{novak2013,delange2017}. Nevertheless, the spin dependence would
not disappear even if we did \cite{delange2017}.


{\change

\section{Convergence of the impurity-model solution with the size of the
  many-body basis}
\label{app:Mconv}

\begin{figure}
\hbox to\linewidth{\hfill%
\includegraphics[width=\figfrac\linewidth]{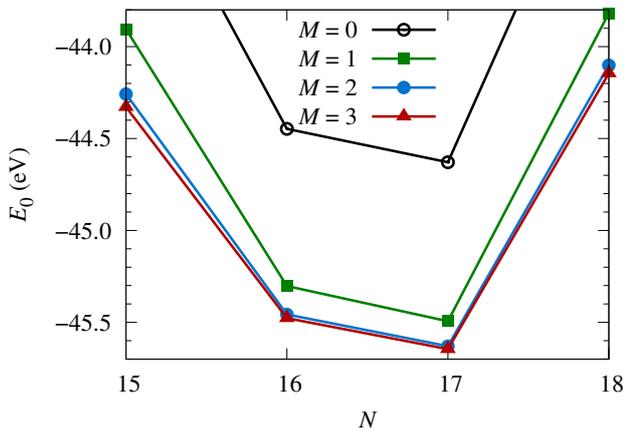}%
\hfill}
\caption{\label{fig:Mconv_Egs}\change The ground-state energy $E_0$ of the
  impurity model at different fillings $N$ computed for increasing
  size of the many-body basis characterized by the cut-off $M$.}
\end{figure}

As indicated in Sec.~\ref{sec:methodGeneral}, we cannot diagonalize
the impurity model in the complete Fock space, only in reduced
subspaces $\mathcal H^{(M)}_N$, defined in
Eq.~\eqref{eq:EDtruncHilbert}, where $N$ is the number of electrons in
the model (its filling) and $M$ is a cut-off parameter. Analyzing the
convergence of the complete DMFT solution with respect to $M$ is
computationally very demanding. Hence, we limit this Appendix to selected
intermediate quantities, evaluation 
of which does not involve computing the self-energy. In particular, we
diagonalize the auxiliary impurity model corresponding to the [210]
ferromagnetic LDA+DMFT solution, obtained for $\Delta_{fd}=35$~meV and
presented in 
Sec.~\ref{sec:results}, for different settings of the cut-off
parameter $M$. The crudest approximation is $M=0$ that does not allow
any hops of electrons between the 5f shell and the bath, and thus
corresponds to the Hubbard-I approximation. The best approximation we
consider is $M=3$, one step better than the setting employed in
the main text.

Figure~\ref{fig:Mconv_Egs} shows the $M$-dependence of the
ground-state energy $E_0$ for fillings $N$ around the overall
grandcanonical ground state which is located at $N=17$. The
differences between the $M=2$ and $M=3$ basis sets are very small
(less than 70~meV), which indicates that $M=2$ is indeed a sensible
choice. The differences are even smaller (less than 30~meV) for energy
gaps $E_0(N\pm1)-E_0(N)$ that determine the positions of the
main peaks in the valence-band spectra like those plotted in
Fig.~\ref{fig:5fDOS}.

\begin{figure}[b!]
\hbox to\linewidth{\hfill%
\includegraphics[width=\figfrac\linewidth]{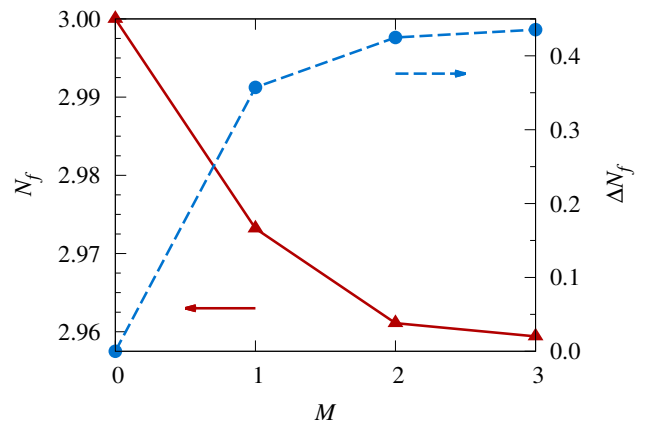}%
\hfill}
\caption{\label{fig:Mconv_occ}\change Convergence of the 5f occupation number
  $N_f$ (red triangles, left axis) and its fluctuation $\Delta N_f$
  (blue circles, right axis) with respect to the basis-set cut-off $M$.}
\end{figure}

Furthermore, we present the 5f occupation number
$N_f=\tr(\hat N_f\,\hat\rho)$, where $\hat\rho$ is the grandcanonical
density matrix of the impurity model, together with its fluctuation
\begin{equation}
\label{eq:fluctNf}
\Delta N_f=\sqrt{
  \tr\bigl(\hat N_f^2\,\hat\rho\bigr) - N_f^2}
\end{equation}
as functions of the cut-off $M$ in Fig.~\ref{fig:Mconv_occ}. Both
these quantities again change very little when $M$ is increased from
$M=2$ to $M=3$, which represents another reassurance that $M=2$ is
good enough.

Note that $N_f$ should be the same number as $n_f$ defined in
Eq.~\eqref{eq:operator_averages} and listed in
Table~\ref{tab:mom_dmft}, which follows from the DMFT embedding
condition. In our DMFT calculations, they are not the same, 
$N_f$ is approximately 0.2 larger than $n_f$, which is a consequence
of the approximate finite impurity model (discrete bath)
being used instead of the exact infinite impurity model (continuous
bath). This is roughly the same discrepancy as we observed earlier when
we applied this method to the ferromagnetic nickel
\cite{kolorenc2012a}. In principle, the situation could be improved by
adding more bath orbitals, but in practice, it is computationally
prohibitive at present.


} 

\bibliography{dft,dmft,codes,xas,UGa2,uranium,pu,AnO,nickel,aim,adatoms,fd_exchange,crystal_field,lanczos}

\end{document}